# Employee Well-being in the Age of AI: Perceptions, Concerns, Behaviors, and Outcomes

Soheila Sadeghi [1]

*Abstract*— The growing integration of Artificial Intelligence (AI) into Human Resources (HR) processes has transformed the way organizations manage recruitment, performance evaluation, and employee engagement. While AI offers numerous advantages—such as improved efficiency, reduced bias, and hyper-personalization—it raises significant concerns about employee well-being, job security, fairness, and transparency. The study examines how AI shapes employee perceptions, job satisfaction, mental health, and retention. Key findings reveal that: (a) while AI can enhance efficiency and reduce bias, it also raises concerns about job security, fairness, and privacy; (b) transparency in AI systems emerges as a critical factor in fostering trust and positive employee attitudes; and (c) AI systems can both support and undermine employee well-being, depending on how they are implemented and perceived. The research introduces an AI-employee well-being Interaction Framework, illustrating how AI influences employee perceptions, behaviors, and outcomes. Organizational strategies, such as (a) clear communication, (b) upskilling programs, and (c) employee involvement in AI implementation, are identified as crucial for mitigating negative impacts and enhancing positive outcomes. The study concludes that the successful integration of AI in HR requires a balanced approach that (a) prioritizes employee well-being, (b) facilitates human-AI collaboration, and (c) ensures ethical and transparent AI practices alongside technological advancement.

*Keywords*— Artificial Intelligence, Human Resources, Employee Well-being, Job Satisfaction, Organizational Support, Transparency in AI

## INTRODUCTION

Integrating Artificial Intelligence (AI) into employee management has significantly reshaped how organizations handle performance evaluations, employee engagement, and overall workforce development [1], [2]. In HR information systems or cloud-based systems, AI can analyze data about employees' learning, personal characteristics, hours worked, and performance measures [3]. While AI offers numerous advantages, including enhanced efficiency and reduced bias, it raises significant concerns—particularly regarding employee well-being. As AI takes on a more significant role in HR functions, understanding its impact on employees' job satisfaction, mental health, and retention has become increasingly crucial [4], [5], [6].

AI-driven HR systems have demonstrated their potential to enhance employee experience and engagement. By providing hyper-personalized experiences and continuous support, these systems improve employee satisfaction and foster loyalty [7]. For instance, AI can significantly reduce the time spent on manual HR tasks, such as performance reviews, while ensuring greater accuracy, fairness, and real-time feedback [8]. However, employees are key stakeholders in the AI-driven workplace, and their perceptions and reactions toward AI can heavily influence the success or failure of its adoption. Interestingly, research indicates that employee attitudes toward AI are often paradoxical—individuals can exhibit both positive and negative attitudes toward AI depending on the context [9]. In this regard, AI transparency emerges as a key factor. For example, employees with positive perceptions of AI tend to view it as an opportunity for growth, resulting in a lower likelihood of leaving their jobs. In contrast, negative perceptions can increase turnover intentions and diminish work engagement, especially in industries such as hospitality [10]. Moreover, a lack of transparency and communication around AI-driven performance evaluations often leads to unpredictability, negatively affecting how employees perceive the system's fairness and their ability to improve based on feedback [11].

Research by [12] confirms that higher awareness of AI and automation negatively affects organizational commitment and career satisfaction, while increasing turnover intentions, depression, and cynicism, particularly among younger employees. Additionally, recent research highlights that when AI is involved in career development decisions, it often lowers employees' perceptions of fairness and satisfaction. It also increases privacy concerns, mainly when AI is the sole decision-maker [13]. While greater transparency can improve perceptions of AI's effectiveness, it may simultaneously cause discomfort, potentially undermining employee trust [14].

This study explores how AI impacts employee well-being, focusing on perceptions, concerns, behaviors, and outcomes. Specifically, it examines how AI shapes job satisfaction, mental well-being, and turnover intentions and what organizations can do to mitigate any adverse effects while enhancing positive outcomes. Moreover, it explores the complex relationship between AI in HR and employee well-being, providing insights for researchers and practitioners. The study introduces an AI-employee well-being interaction framework highlighting how AI influences employee perceptions, behaviors, and outcomes. This framework offers a practical guide for understanding AI's impact on well-being and shaping future research and strategies.

1. Soheila Sadeghi is with the University of the Incarnate Word, San Antonio, TX 78209, USA (corresponding author, e-mail: ssadeghi@student.uiwtx.edu).

Employee perceptions of AI vary widely, often shaped by specific use cases in HR processes. Some employees view AI as a tool that enhances efficiency and reduces biases in decision-making [15], while others express concerns about fairness, transparency, and potential dehumanization in their interactions [16]. Employees with positive attitudes toward "Intelligent Automation" emphasize the convenience and rationality AI can offer [9]. In contrast, those with negative "No Human Interaction" attitudes prefer human engagement, affecting their openness to AI technologies.

However, even when AI is implemented to support decision-making, research shows it may not eliminate adverse employee reactions. Employees often perceive AI-supported decisions as less fair than those made solely by humans [13]. When employees are unable to understand how AI evaluates their performance, they become frustrated and helpless, leading to growing distrust [11]. This lack of understanding can create a perception that AI lacks human qualities, such as empathy, which diminishes perceptions of fairness, trust, and appropriateness—especially in HR contexts like hiring, firing, and performance evaluations [17].

Additionally, an employee's perception of AI is often influenced by their understanding of how these systems operate. When AI decision-making lacks transparency, it can breed mistrust and negative views about fairness [18]. While transparency is critical for fostering employee trust, it is not always sufficient. Studies indicate that even when AI decisions are transparent if perceived as intrusive or unfair, employee satisfaction may decline, reinforcing negative views about AI's role in HR [8], [13].

On the other hand, when AI is implemented with the right level of transparency and personalization, it can enhance trust and engagement. For example, integrating AI systems into HR ecosystems has created positive experiences, offering hyper-personalized responses and support that improve employee loyalty [7]. However, transparency can have a dual effect: while it improves perceived effectiveness, it may also raise discomfort, resulting in a complex interplay between trust-building and trust-eroding mechanisms [14].

As AI continues to take on critical roles, such as evaluating employee performance and scheduling tasks, many workers report losing control over their work, leading to increased anxiety and stress [16]. For instance, research on AI and robotics awareness among hotel employees found positive perceptions of AI as a job efficiency tool to improve work engagement. In contrast, fears of job displacement reduce engagement and increase turnover intentions [10]. When employees experience unpredictable changes in AI-driven performance evaluations, such as fluctuating scores without clear explanations, their stress levels rise, highlighting the need for better communication and transparency in AI implementation [11].

## I. KEY CONCERNS ON AI AND EMPLOYEE WELL-BEING

### A. Job Security Concerns

A central concern for employees regarding AI is job security. Many workers fear AI will eventually replace them, particularly in roles where routine tasks are automated [2], [19]. These concerns can have a profound impact on psychological well-being, often manifesting as anxiety and insecurity. For example, research shows that AI and robotics awareness among hotel employees significantly increase turnover intentions, driven by concerns over job displacement [20]. In sectors like hospitality, AI's efficiency in automating tedious HR tasks further exacerbates these fears, as employees worry about losing their jobs to machines [8]. As [9] highlights that such fears are often rooted in job displacement concerns and the broader uncertainty surrounding human-machine interfaces. This uncertainty intensifies when employees feel powerless or do not understand how AI systems function.

An example of this can be seen in freelancers on platforms like TalentFinder, who express heightened paranoia about arbitrary score fluctuations—leaving them uncertain about their job security [11]. When employees feel they have no recourse or control over how AI systems evaluate them, these concerns escalate, further amplifying feelings of vulnerability and instability.

### B. Fairness and Transparency

The perceived fairness of AI systems significantly shapes employee attitudes. AI-driven decisions, particularly in performance evaluations, are often seen as less fair than human evaluations due to the reductionist nature of algorithms, which may overlook qualitative aspects of performance [21]. While AI-based performance evaluation systems can improve fairness and accuracy by reducing human bias, the lack of empathy and personalization in AI-driven decisions can diminish justice perceptions [8].

A critical issue in fairness is the opacity of AI decision-making processes. The lack of transparency in AI evaluation systems makes employees feel uninformed and powerless to improve, undermining their trust in the system and reducing job satisfaction [11]. However, when AI systems are well-designed with transparent decision-making processes, they can enhance the employee experience by enhancing trust and promoting higher engagement [7]. Transparency is essential for improving perceptions of fairness.

Studies show that when employees view AI as a tool to enhance their career prospects, they tend to perceive higher levels of organizational support, which helps mitigate the negative effects of AI on job security and fairness [10]. Employees with more balanced "Intelligent Automation" attitudes often prioritize AI's practical benefits over concerns about fairness, focusing on the positive impact on their work rather than on its potential downsides [9].

However, transparency is not without its challenges. While it can improve fairness perceptions, too much transparency can sometimes create discomfort and negatively impact trust [14]. For instance, when AI is involved in career development decisions, it diminishes fairness perceptions. It raises privacy concerns, especially when external data sources are combined with internal data [13] in environments where employees cannot understand the criteria behind AI-driven performance scores [11].

Additionally, informing employees about how AI systems make decisions leads to more positive views of those decisions [18]. The right balance of transparency is key to building trust and improving fairness in AI-driven HR processes.

### C. Lack of Control

AI-driven systems, especially scheduling and performance management, can give employees less control over their work [22]. This loss of control contributes to feelings of dehumanization, where employees feel treated more like data points than individuals [23]. Workers report feeling at the mercy of unpredictable algorithms, reducing their autonomy and heightening stress and anxiety [11].

While AI-driven performance evaluations can increase perceptions of objectivity and fairness, some employees feel a loss of control over their work, which leads to stress and dissatisfaction [8]. Those with "No Human Interaction" attitudes are particularly resistant to AI systems that remove personal interaction, as they prefer decision-making processes that involve emotional intelligence and human empathy [9].

Many employees view AI as lacking the emotional intelligence necessary for meaningful human interactions, which can diminish trust and increase feelings of disrespect [17]. In organizations where AI is implemented without addressing these concerns, there tend to be lower work engagement and higher turnover intentions [10]. Feelings of dissatisfaction and dehumanization rise when AI alone makes decisions, reducing engagement and increasing turnover [3]. Employees often prefer human involvement in decision-making because it allows for empathy and understanding—qualities AI systems inherently lack [16].

However, some organizations, like BigTech, have effectively mitigated these concerns by involving employees in designing and implementing AI systems. Doing so enhances trust and engagement, offering more personalized experiences that make employees feel heard and valued [7]. This approach addresses the issue of control and helps employees feel more connected to technology, fostering a healthier work environment.

### D. Mental Health and Privacy Concerns

AI systems that track employee activities or use data to monitor performance can exacerbate privacy concerns and increase stress [19]. Employees worry that constant surveillance could infringe on their personal space, leading to a decline in mental well-being. This is particularly relevant when AI-driven decisions are made using internal and external data sources, as the increased data volume often intensifies privacy concerns, leaving employees feeling intruded upon [13]. These concerns are heightened when employees do not fully understand how AI systems are being used or what data is being collected [18]. However, when AI is employed to enhance employee well-being through health and wellness applications that promote mental and physical wellness, AI can contribute positively to employee satisfaction and reduce stress [7]. This is exacerbated when AI systems are not transparent, where employees are left uncertain about what data is being collected and how it affects their job prospects. This lack of clarity can result in increased paranoia and anxiety, significantly impacting mental well-being [11].

## II. THE IMPACT ON JOB SATISFACTION AND RETENTION

### A. Job Satisfaction

AI's impact on job satisfaction is complex. On the one hand, AI can improve job satisfaction by automating routine tasks, allowing employees to focus on more meaningful work [15]. For example, employees reported feeling anxious about how the algorithm evaluated their work. This decreases job satisfaction when they don't understand the evaluation criteria or how to improve their scores [11]. However, job satisfaction tends to increase when employees perceive AI systems as fair and supportive, especially in performance evaluations [8].

Attitudes towards AI also play a role in job satisfaction. Employees with "No Human Interaction" attitudes may experience reduced job satisfaction due to concerns about AI replacing human interactions. In contrast, those with "Intelligent Automation" attitudes may appreciate AI's convenience and focus on efficiency [9]. Research shows that job satisfaction decreases when AI-driven career development processes are perceived as unfair or intrusive, particularly when decisions are made without human involvement [13]. Similarly, when AI reduces fairness or transparency in processes like performance management, employees report lower job satisfaction [21].

Unexplained or unpredictable changes in AI-driven evaluations can further erode job satisfaction, as employees feel unsupported and disempowered [11]. Studies suggest that perceived organizational support can moderate the relationship between AI awareness and job satisfaction, reducing turnover intentions and improving overall satisfaction [10]. In fact, when employees feel their organization supports them in understanding and navigating AI systems, it helps address concerns, enhancing job satisfaction [20]. AI-driven decisions that result in positive outcomes are linked to higher levels of job satisfaction and trust, while negative outcomes—regardless of whether a human or AI made the decision—can lead to diminished perceptions of fairness [17]. For AI to positively impact job satisfaction, it must be perceived as a tool that enhances employee contributions and workplace autonomy rather than diminishing them [16]. When AI systems offer transparency and personalization, as seen in advanced HR ecosystems, employees report improved job satisfaction because they feel their contributions are valued, and their autonomy is respected [7]. However, while transparency in AI can improve perceived effectiveness and satisfaction, it also carries the risk of creating discomfort, which could undermine these benefits [14]. Achieving the right balance between transparency and personalization is key to ensuring AI systems positively contribute to job satisfaction.

*B. Employee Retention*

Negative perceptions of AI, particularly regarding job security and fairness, can lead to higher turnover intentions [24]. While AI-driven systems that reduce bias and improve decision-making in HR can help lower turnover by offering more objective and personalized evaluations [8], the lack of transparency in AI systems remains a significant retention challenge. For instance, workers who experienced unexplained score drops were more likely to leave the platform, illustrating how unpredictable AI-driven systems can drive disengagement [11]. Employees who perceive AI evaluations as arbitrary and unpredictable often disengage from their work, leading to higher intentions to leave. This highlights the critical need for greater transparency in AI performance systems [11]. In sectors like hospitality, employees with negative views of AI, especially those who see it as a threat to their job security, are more likely to leave their organizations [10]. As [9] highlights the importance of managing these perceptions, as employees who feel replaced or dehumanized by AI are more inclined to exit their jobs. Research further shows that turnover intentions rise significantly when AI decisions are perceived as intrusive or unfair, with privacy concerns as a key factor in this process [13].

On the other hand, AI systems that enhance the employee experience through hyper-personalization and continuous feedback have been found to reduce turnover, as they create a more engaging and supportive work environment [7]. In workplaces with competitive psychological climates, the relationship between AI awareness and turnover intentions intensifies, particularly when employees feel stressed or pressured by competition [20].

When employees feel that AI systems are unfair or threaten their job security, they are more likely to consider leaving their organizations [19]. Conversely, employees are more likely to stay with their organizations when AI is perceived as fair and when positive outcomes are achieved—regardless of whether they are determined by a human or AI system [17]. Moreover, when AI is implemented with transparency and seen as complementing human work rather than replacing it, it can strengthen organizational commitment and reduce turnover intentions [18]. In this sense, effectively managing AI's role in the workplace is crucial for improving employee retention.

To further illustrate the various dimensions of AI's impact on employee well-being, table I outlines both positive and negative impacts of AI on key employee concerns such as perceptions, job satisfaction, mental well-being, fairness, and retention. The table also highlights actionable organizational strategies to mitigate negative outcomes and enhance positive ones.

## III. THE ROLE OF ORGANIZATIONAL SUPPORT

*A. Transparency and Communication*

As Table I outlines, organizations play a crucial role in mitigating negative employee perceptions of AI by ensuring transparency and clear communication about its use. The lack of clear communication about how the AI evaluation system operates has led to confusion and anxiety among employees, who feel powerless to improve their standing or avoid penalties [11]. Providing detailed explanations of how AI systems function and how decisions are made can significantly increase trust while reducing concerns about fairness and job security [25].

Involving employees in the development of AI systems, as demonstrated by organizations like BigTech, further boosts trust by addressing employee concerns and respecting their roles in an AI-driven workplace [7]. When employees understand how AI systems work and feel they have some influence over the outcomes, they are more likely to engage positively with these systems [11]. This sense of understanding is critical, as employees need to feel that AI is supporting—not replacing—them in their work environment [18].

Trust in decision-making is key to enhancing perceptions of fairness, or "interactional justice," especially when AI systems are implemented appropriately, and outcomes are communicated transparently [17]. Organizational support is essential to alleviating fears of job loss due to AI. Employees who feel supported by their organization are less likely to leave and more likely to engage positively with AI systems [10]. Moreover, ensuring transparency, communication, and employee involvement in AI initiatives builds trust and fosters a more engaged and stable workforce. Table I summarizes the key challenges AI presents in the workplace, their impact on employee well-being, and recommended strategies for addressing them. It serves as a practical guide for HR managers and leaders to implement AI in ways that promote engagement and minimize negative effects. Table I summarizes the key challenges AI presents in the workplace, their impact on employee well-being, and recommended strategies for addressing them. It serves as a practical guide for HR managers and leaders to implement AI in ways that promote engagement and minimize negative effects.

TABLE I
KEY AI CHALLENGES, EMPLOYEE WELL-BEING IMPACTS, AND ORGANIZATIONAL STRATEGIES

| Category | Key Issues/Concerns | Implications for Employee Well-being | Recommended Organizational Strategies |
|---|---|---|---|
| Job Security | Fear of job replacement, automation of routine tasks | Increased anxiety, stress, and turnover intentions | Upskilling and reskilling programs; transparency on AI's role in complementing human work |
| Fairness and Transparency | Opaque AI decision-making, lack of communication | Reduced trust, job satisfaction, and perceived fairness | Provide clear communication on AI decision-making processes; Involve employees in AI implementation |
| Control and Autonomy | AI-driven performance evaluations and scheduling limiting employee control | Feelings of dehumanization, decreased work engagement | Involve employees in AI system design; ensure AI complements rather than replaces human decision-making |
| Mental Health and Privacy | AI monitoring employee activities, lack of transparency around data use | Increased stress, paranoia, and reduced mental well-being | Implement privacy safeguards; Communicate how data is collected and used |
| Job Satisfaction | AI automating tasks, unpredictable AI-driven evaluations | Increased job satisfaction when AI is supportive; decreased when AI decisions are unclear | Personalize AI systems to enhance job satisfaction; Ensure transparency and fairness in AI-driven processes |
| Employee Retention | Negative perceptions of AI regarding fairness and job security | Higher turnover intentions, disengagement from work | Promote fairness and transparency in AI systems; Offer continuous feedback and skill development programs |
| Trust and Transparency | Lack of AI system transparency leading to mistrust | Erosion of trust, increased employee turnover | Ensure AI systems are transparent and communicate decisions effectively to employees |
| Positive AI Perceptions | AI as a tool for efficiency, convenience, career development | Higher productivity, greater engagement, lower turnover intentions | Foster "Intelligent Automation" attitudes; Highlight AI's role in enhancing employee contributions |
| Negative AI Perceptions | AI seen as intrusive, dehumanizing, or unfair | Increased absenteeism, reduced engagement, turnover intentions | Manage AI perceptions through employee involvement, transparency, and human-AI collaboration |
| Organizational Support | Lack of support for employees in adapting to AI | Higher stress, lower job satisfaction, higher turnover intentions | Offer strong organizational support through upskilling programs, clear communication, and regular feedback mechanisms |

### B. Upskilling and Reskilling Programs

Offering employees opportunities to learn new skills and adapt to working alongside AI can significantly alleviate job security concerns [19]. Personalized training programs that focus on enhancing employees' technological skills reduce turnover intentions and improve engagement with AI systems [10]. Organizations can adopt a more balanced and positive view of AI by creating an environment where AI complements human work rather than replacing it [9].

Reskilling programs tailored to align with AI-driven transformations in HR, such as competency development and career growth initiatives, have been shown to boost employee confidence and reduce turnover intentions [7]. Employees who feel empowered by these programs are more likely to see AI as an opportunity for growth rather than a threat. Additionally, these initiatives can help employees feel more secure in their roles, leading to higher job satisfaction and lower turnover intentions [24].

Ultimately, offering upskilling and reskilling programs is critical in helping employees embrace AI-driven changes, ensuring they feel supported and capable in the evolving workplace.

### C. Involving Employees in AI Rollouts

Involving employees in designing and implementing AI systems can significantly improve their perceptions of the technology [21]. By actively soliciting feedback and ensuring that employees feel heard, organizations can reduce resistance to AI while fostering a collaborative culture where humans and AI work together.

## IV. BEHAVIORAL OUTCOMES LINKED TO AI PERCEPTIONS

### A. Positive Behaviors

Employees with positive perceptions of AI are more likely to engage with the technology and demonstrate higher productivity [15]. Employees who view AI as a tool for career development tend to show greater work engagement and lower turnover intentions, especially in industries where technology is seen as a support system rather than a replacement [10]. In contrast, platforms where AI systems lack transparency can cause even high-performing workers to disengage, as uncertainty about maintaining or improving their performance creates frustration [11].

Employees with "Intelligent Automation" (IA) attitudes are more open to adopting AI technologies, particularly when they see practical benefits such as increased efficiency and convenience [9]. AI systems that provide personalized, real-time feedback and continuous support, as seen in advanced HR ecosystems, foster positive workplace behaviors, increased collaboration, and higher engagement [7]. Employees often welcome AI that reduces workloads and enhances efficiency, resulting in improved workplace behaviors like greater collaboration and engagement [26]. Moreover, when AI-driven decisions produce favorable outcomes, employees are more likely to trust the technology and experience less dehumanization [17].

## B. Negative Behaviors

Negative perceptions of AI, especially concerning fairness and job security, can lead to leaving behaviors such as increased absenteeism and reduced engagement [19]. For example, freelancers expressed frustration and disempowerment when AI-driven evaluations caused unpredictable score drops, leading to disengagement and higher turnover intentions [11]. Employees who feel threatened by AI, particularly those who view it as a hindrance rather than a tool, are more likely to disengage from their work and consider leaving their organizations [10].

Employees with "No Human Interaction" attitudes are more resistant to adopting AI, especially when they believe it diminishes personal interactions and emotional intelligence [9]. AI systems that fail to address concerns about fairness and transparency can lead to disengagement and mistrust, ultimately affecting organizational outcomes [7]. Distrust of AI also makes employees less likely to adopt new technologies and can create resistance, reducing overall productivity and fostering conflict within teams [23]. When AI decisions are perceived as unfair or dehumanizing, employees experience lower interactional justice and diminished trust and are more likely to disengage from their work [17].

In both cases, addressing employee concerns and ensuring transparency in AI-driven systems are essential for fostering positive perceptions and minimizing negative outcomes.

## V. AI-EMPLOYEE Well-being Interaction Framework

The AI-Employee Well-being Interaction Framework (Fig. 1) illustrates how AI integration into HR processes, such as recruitment, performance management, employee engagement, and scheduling, influences employee perceptions and behaviors. These AI-driven processes lead to Employee Perceptions and Concerns, both positive (e.g., efficiency improvements, reduced bias) and negative (e.g., job security, fairness, and privacy concerns).

Moderating Factors—including Transparency and Communication, Organizational Support, and Employee Attitudes—amplify or mitigate these concerns, shaping employee reactions to AI. These perceptions drive various Employee Behaviors and Outcomes, with positive outcomes like increased job satisfaction and lower turnover intentions and negative outcomes such as increased stress and disengagement.

The Feedback Loop shows how these outcomes influence future perceptions, creating a continuous cycle between AI, employee perceptions, and well-being. This framework highlights key areas for intervention to maximize positive outcomes and minimize negative effects. The framework also serves as a model for future research. Future studies could focus on longitudinal studies, contrastive analyses, and bibliometric analysis to explore the impact of AI characteristics on HR outcomes in the future [27].

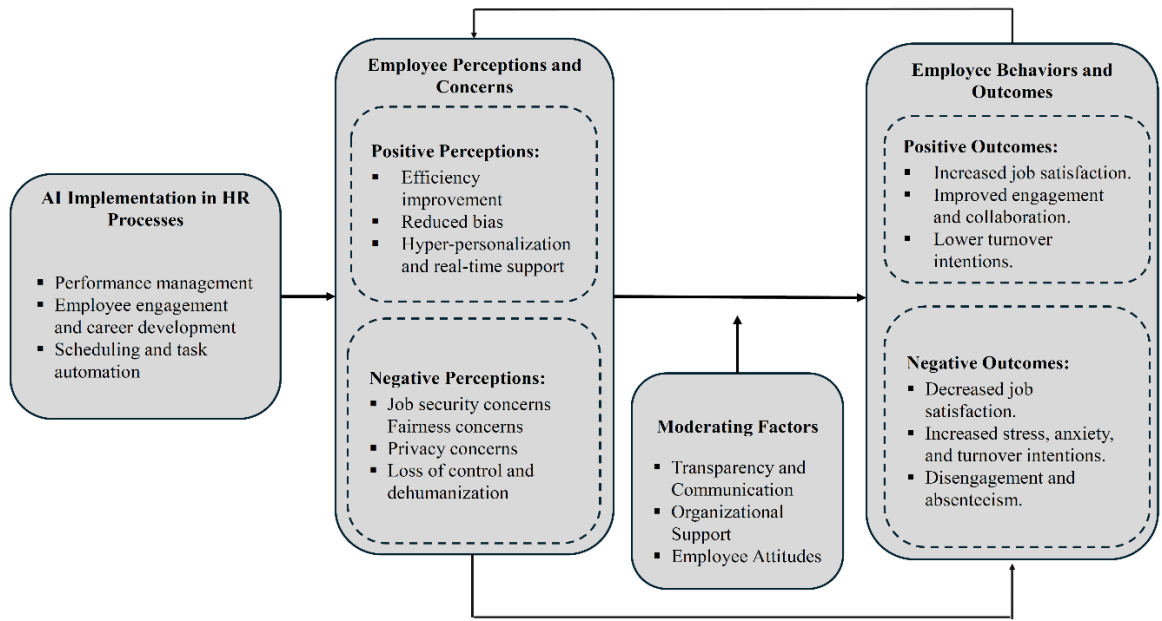

Fig. 1: AI-Employee Well-being Interaction Framework

## VI. FUTURE OF AI AND EMPLOYEE WELL-BEING

As AI evolves, organizations must ensure its use enhances rather than hinders employee well-being. The opaque nature of AI-driven evaluations has led to significant stress and disengagement, highlighting the need for transparency and employee involvement to mitigate these issues [11]. Integrating AI into HR processes with transparency, personalization, and employee involvement can lead to enhanced employee experiences, increased job satisfaction, and improved retention [7]. These findings, aligned across different domains, underscore a universal principle: transparency builds trust, whether in financial decisions or organizational change. In financial contexts, transparency in portfolio selection demonstrates trust through alignment of investment decisions with individual risk preferences [28]. Similarly, clear and transparent communication about organizational changes has been shown to lead to more positive market reactions [29].

Organizations must carefully balance AI's benefits with its challenges. Perceptions of AI as a career-enhancing tool can reduce turnover intentions and increase engagement, while negative views of AI as a threat can have the opposite effect [10]. Managing employees' paradoxical attitudes toward AI is crucial, as some may exhibit positive and negative reactions depending on the context [9]. Future advancements in AI should focus on increasing transparency, improving fairness, and fostering human-AI collaboration [16].

However, organizations must be cautious. While greater transparency can boost perceived effectiveness, it can also increase discomfort, negatively impacting trust and employee well-being [14]. Transparency alone is not always enough to improve trust—how decisions are made and communicated plays a critical role. To ensure AI enhances employee well-being, organizations must prioritize its ethical use, ensuring employees feel valued and supported in a technology-driven environment [21].

For AI to truly enhance employee well-being, it must be implemented carefully, considering fairness, respect, and the way decisions are communicated and perceived [17]. By doing so, companies can create workplaces where AI and employees coexist harmoniously, fostering innovation and well-being.

## IX. PRACTICAL IMPLICATIONS

The findings of this study have significant implications for HR managers and organizational leaders implementing AI in their workforce management strategies. While AI offers efficiency and cost-saving benefits, a thoughtful approach is essential. Transparency about AI systems and their decision-making processes is key to easing concerns and boosting satisfaction. Maintaining human oversight of critical decisions helps preserve trust, while privacy concerns should be addressed by informing employees about data usage. Involving employees in AI implementation and providing training can increase engagement and position AI as a tool for growth. Regular audits for bias and strong feedback mechanisms ensure ethical AI use. A gradual, adaptive approach allows organizations to harness AI's potential while maintaining a human-centric work environment.

## X. CONCLUSION

The rise of AI in HR has brought about significant changes in how employees experience their work, and it presents both opportunities and challenges for their well-being. This review highlights several important insights: a) employees' views on AI are mixed, with some seeing it as a way to make their work more efficient, while others worry it threatens job security and reduces personal connections; b) when AI systems are not transparent, employees often feel distrustful and perceive them as unfair and c) AI-driven processes can lead to feelings of losing control over one's work, which can cause stress and dissatisfaction. Additionally, the use of AI to monitor employees or make decisions raises privacy issues, which can add to stress and anxiety.

However, using AI thoughtfully can boost job satisfaction by taking over routine tasks and offering personalized experiences. On the other hand, negative perceptions of AI, especially when it comes to job security and fairness, can increase employees' desire to leave their jobs. Organizations play a vital role in shaping how employees feel about AI and key strategies for success include providing transparency, clear communication, opportunities for skill development, and involving employees in the rollout of AI systems. Behaviorally, d) employees who have a positive view of AI tend to be more engaged and productive, while e) those with negative views may withdraw and be less involved in their work.

As AI becomes more integrated into the workplace, organizations must carefully balance the benefits and challenges it brings. To succeed, they should focus on f) improving transparency and making AI decisions easier to understand; g) ensuring fairness and ethical use of AI in HR practices; h) promoting collaboration between humans and AI, rather than replacing employees with technology; i) providing ongoing support, training, and skill development; and j) involving employees in how AI is designed and used. When organizations prioritize these aspects, they create a work environment where AI supports, rather than disrupts, employee well-being. The ultimate goal is to build workplaces where AI and employees work together harmoniously, driving innovation while maintaining a focus on human needs, autonomy, and job satisfaction.

As AI continues to advance, its impact on employee well-being will remain a key concern. Companies that can navigate this new landscape successfully will not only increase employee satisfaction and retention but also strengthen their ability to attract and keep top talent in an increasingly AI-driven world. The challenge lies in balancing technology and human-centered approaches, ensuring that AI in the workplace enhances the quality of life for employees rather than diminishing it.

APPENDIX

N.A.

ACKNOWLEDGMENT

N.A